\documentclass[11pt,preprint]{aastex}

\def\myplot#1#2#3#4#5#6#7{\centering \leavevmode
\vbox to#2{\rule{0pt}{#2}}
\includegraphics{#1}}

\begin{document}

\title{Infrared Interferometric Observations of Young Stellar Objects \footnotemark[1]}

\footnotetext[1]{to be published in the {\it Astrophysical Journal}}
\author{R.L. Akeson\altaffilmark{2,3}, D.R. Ciardi\altaffilmark{4},
G.T. van~Belle\altaffilmark{3}, M.J. Creech-Eakman\altaffilmark{3} and
E.A. Lada\altaffilmark{4}}

\altaffiltext{2}{Infrared Processing and Analysis Center, California Institute of Technology MS 100-22, Pasadena, CA, 91125}

\altaffiltext{3}{Jet Propulsion Laboratory, MS 171-113, 4800 Oak Grove, Pasadena, CA 91109}

\altaffiltext{4}{University of Florida, 211 Bryant Space Sciences Bldg, Gainesville, FL 32611}

\begin{abstract}

We present infrared observations of four young stellar objects
using the Palomar Testbed Interferometer (PTI).  For three of the
sources, T Tau, MWC 147 and SU Aur, the 2.2 $\mu$m emission is
resolved at PTI's nominal fringe spacing of 4 milliarcsec (mas), while the
emission region of AB Aur is over-resolved on this scale.  We fit the
observations with simple circumstellar material distributions and compare our
data to the predictions of accretion disk models inferred from
spectral energy distributions.  We find that the infrared emission
region is tenths of AU in size for T Tau and SU Aur and $\sim$1 AU for
MWC~147.

\end{abstract}


\section{Introduction}

Observational evidence for circumstellar material around most young
stellar objects (YSOs) includes infrared emission in excess of that
expected from the stellar photosphere, broad forbidden line profiles,
and emission at millimeter wavelengths.  Although the dust column
density is inferred to be quite high, the sources are often optically
visible, implying a geometrically flat distribution of the material.
A disk morphology is also predicted by star formation theories as a
consequence of conservation of angular momentum.  Evidence for
circumstellar disks has been observed around sources with a range of
masses, from near solar (T Tauri stars) to greater than 10 M$_{\odot}$
(Herbig Ae/Be stars) (see e.g. \citet{mun00,nat00}).  Disks not only
provide a conduit for material to accrete onto the central star, but
are also a reservoir of material from which a potential planetary
system might form.

The structure of YSO circumstellar disks has been studied using
spectral energy distributions (SED), spectral line profiles and
imaging at infrared and (sub)-millimeter wavelengths.  The dust
continuum emission from disks around several T Tauri sources has been
resolved at millimeter wavelengths (see review by \citet{wil00}).
These observations are sensitive to emission from cooler dust and
provide spatial information on size scales of several 10's of AU. The
disk physical properties on much smaller scales ($<$ few AU) are
generally inferred through examination of the spectral line shapes and
modeling of the SED.  Unresolved issues regarding the inner disk structure
include the possible existence of inner disk holes
(e.g. \citet{hil92}) and the validity of simple power law scalings
to describe globally the temperature and density profiles 
of the disk.  Characterizing the physical properties of the inner disk
is important for theories of hydrodynamic disk winds and for understanding
the initial conditions of planet formation.

Infrared interferometry provides a method to directly observe the
inner disk.  To date, only a few YSOs have been observed using this
technique (e.g. FU Ori: \citet{mal98} and AB Aur: \citet{mil99a}).
Here we present K-band long baseline interferometric observations of
four YSOs: T~Tau, SU~Aur, AB~Aur, and MWC~147.  PTI has a fringe spacing
of $\sim$4 mas, corresponding to 0.6 AU at the distance of
Taurus-Aurigae (140 pc) and to 3 AU at the distance of MWC~147 (800
pc).

\section{Observations and data reduction}

Observations were made in the K band at PTI, which is described by
\citet{col99}.  The data were obtained between September and December
1999.  For each source, the number of nights, the total number of records
(each of which contains 25 seconds of data) and the calibrators used are
given in Table 1.  The data were calibrated using the standard method
described in \citet{bod98}.  A synthetic wideband channel is formed
from the five spectrometer channels.  The system visibility is
measured with respect to the calibrators.  The calibrator sizes were
estimated using a blackbody fit to photometric data from the
literature and were confirmed to be internally consistent when two or
more calibrators were observed in a given night, which occurred on
most nights.  The calibrators were chosen by their proximity to the
sources and for their small angular size, minimizing systematic errors
in deriving the system visibility.  All calibrators used in this
reduction have angular diameters $<0.8$ mas and were assigned
uncertainties of 0.1 mas, except for HD 46709 ($\theta = 1.8 \pm 0.2\
{\rm mas}$).  The data are presented in normalized squared visibility,
which is an unbiased quantity.  The averaged squared visibility and
error are given for each source in Table 1.  The uncertainties for the
calibrated visibilities are a combination of the calibrator size
uncertainty and the internal scatter in the data.

\section{Models}

Of the four objects discussed in this paper, three were resolved and
the fourth was over-resolved (no fringes were detected).  In this
section we first detail the models and then discuss each source
separately in the following section.  The models fit to the data are a
uniform brightness profile, a Gaussian brightness profile, and a
binary companion.  We also compare the data to accretion disk models.
The uniform and Gaussian profiles are presented as simple
geometric distributions which can be used as size scale estimators.
At the distances to these systems (140 to 800 pc) the central star is
unresolved ($\theta < 0.1$ mas).  For the uniform and Gaussian
distributions and the accretion disk models, a stellar component has
been included as an unresolved source with the appropriate flux ratio.
This ratio was determined by subtracting the photospheric flux from a
star of the appropriate spectral type from the total K band flux.  For
these objects, the infrared emission on milliarcsec scales is
dominated by thermal emission and scattering can be neglected
\citep{mal95}.

\subsection{Uniform and Gaussian profiles}

Two of the simplest geometric models which can be used to describe the
circumstellar material distribution are a uniform brightness
profile and a Gaussian profile.  For a face-on Gaussian or
uniform profile, the predicted visibility is simply a function of the
projected baseline.  For the uniform profile, the squared visibility is
\begin{equation}
V^2 = \left[ \frac{2J_1(\pi \theta B_p / \lambda)}{\pi \theta B_p / \lambda} \right]^2
\end{equation}
where $J_1$ is a Bessel function, $\theta$ is the diameter,
and $B_p$ is the projected baseline.  For a Gaussian profile
\begin{equation}
V^2 = \left( \exp \left[-\frac{\pi^2}{\ln 2}\left(\frac{D}{2}\right)^2 \frac{B_p^2}{\lambda}\right] \right)^2
\end{equation}
where $D$ is the FWHM.

For an inclined profile, the visibility is also a function of hour
angle.  As none of the sources show definitive visibility structure
with hour angle, we limit ourselves to the simple face-on case.  The
observations cover hour angle ranges of -2 to 2 hours for T Tau, -2 to
0 hours for SU Aur and -3 to 1 hours for MWC~147.  Although other
inclinations and position angles are not necessarily excluded by the
data, we note that inclination angles near edge-on would produce
significant visibility variations with hour angle, which are not seen.
The best fit uniform and Gaussian profile sizes for each source are
given in Table 1.

\subsection{Binary companion}

A reduction from unity visibility can be produced by a binary companion.
If both components are individually unresolved, the visibility is given by
\begin{equation}
V^2 = \frac{1 + R^2 + 2R \cos[(2\pi /\lambda)\bf{B} \cdot \bf{s}]}{(1 + R)^2}
\end{equation}
where $R$ is the flux ratio, $\bf{B}$ is the baseline vector and
$\bf{s}$ (in radians) is the binary angular separation vector.  To
test if the measured visibilities are consistent with a binary, a grid
of binary parameters was formed and model visibilities were calculated and
compared to data binned by projected baseline.  The binary parameter space
considered contained primary/secondary flux ratios ($R$) from 1 to 30
and separations ($s$) up to 100 mas, which corresponds to the coherence
length of the spectral channels.  Binary companions beyond this separation
with sufficient magnitude to affect the measured visibility
have been ruled out by speckle or adaptive optics observations
(T Tau and SU Aur, \citet{ghe93}; MWC147, \citet{cor98}).

\subsection{Accretion disk}

Emission from an accretion disk is one of the leading explanations for
the infrared excess and other observed features of T Tauri stars and
has also been proposed for Herbig Ae/Be stars.  One common method of
describing the physical properties of the accretion disk is to
parameterize the temperature ($T$) and surface density ($\Sigma$) as
power-law functions of the radius ($T \propto r^{-q}, \Sigma \propto
r^{-p}$) and the dust opacity as a power-law function of wavelength
($\kappa \propto \lambda^{\alpha}$).  At 2.2 $\mu$m the disk is
optically thick and the emission profile at a given radius depends on
the temperature distribution \citep{bec90}.  As we have only sampled
one spatial scale in the disk, we will use accretion disk models from
the literature, where the SED or millimeter imaging has been used to
determine the disk parameters.

\section{Results}

\subsection{T Tau}
\label{ttau_sec}

T Tauri, one of the best-studied YSOs, has an infrared
companion, T Tau S, 0\farcs 7 to the south, which is optically obscured.  Both
components have a near-infrared excess, suggestive of circumstellar
material.  Recent observations \citep{kor00} have revealed that
T Tau S is also a binary.
The millimeter wave flux is dominated by material
surrounding T Tau N and is consistent with circumstellar disk models
with an outer radius of 40 AU \citep{ake98}.

At K band, T~Tau~N is the component with higher flux
and thus, it contributes most to the
measured visibility.  The binary separation is sufficiently large such
that the fringe envelopes of the two sources do not overlap and thus T
Tau S does not contribute any coherent flux.  However, it is within
the field of view of the star and fringe trackers and therefore
contributes incoherent flux, reducing the observed $V^2$ on T Tau N by
$R^2/(1 + R)^2$, where $R$ is the flux ratio between T Tau N and S.
For these observations the ratio has two components,
the intrinsic K band flux ratio of the system and an instrumental flux
ratio introduced by an optical fiber, both of which are discussed
below.

At PTI, the angle tracking passband is 0.7-1.0 $\mu$m.  T Tau N is
brighter than T Tau S in the visible, \citet{sta98} measured a ratio
of $>$2000 in the I band; thus the field will be centered on T Tau N.
As the optical path for the spectrometer channels includes a fiber
with 1\arcsec\ FWHM, the flux contribution from T Tau S is reduced.
The relative coupling of the N/S flux was calculated in the following
way.  An Airy pattern given by the diffraction limit of
1\farcs2 was convolved with a 1\arcsec\ Gaussian representing the
seeing.  We then use a matched filter of a Gaussian with 1\arcsec\
FWHM for the fiber, which is centered on T Tau N.  The errors were
conservatively estimated by assuming the fiber FWHM to have an error of
50\%.  The resulting fiber coupling ratio (N/S) is 1.50 $\pm$0.2.

The total flux and flux ratio for the T Tau system varies on time scales of
weeks to months (e.g. \citet{skr96}) and so roughly contemporaneous flux
ratios are necessary.  T. Beck (private communication) measured a N-S
flux ratio of 1.96$\pm$0.19 on 31 Oct 1999.  The last four nights of
our data were taken on 20 and 26 Oct and 2 and 3 Nov 1999.  In the
analysis below, we assume the flux ratio measured by Beck is valid for
these 4 nights and use only these data.
  
We note that we have no direct way of knowing that the detected
fringes arise from T Tau N instead of T Tau S.  An alternate
explanation for the detection is that the fringes arise from T Tau S.
If this were the case, the expected visibility with T Tau S unresolved
is $V^2$=0.07, much lower than that observed.  The low visibility is
due to the flux and fiber ratios still favoring T Tau N and would be
even lower if T Tau S were resolved.  The stability of the fringe
tracking and the consistency of the night to night measurements
strongly suggest that only 1 of the binary components is detected, and
given the above argument, we deduce that T Tau N is the detected
component.

To correct for T~Tau~S, the visibilities were divided by $R^2/(1 +
R)^2$ before modeling was performed, where $R$ represents the
correction for both the intrinsic flux ratio and the coupling effect
described above and is 2.94 $\pm$0.48.  For the stellar parameters
estimated by \citet{ghe91} using optical photometry, the total to
stellar flux ratio at K is 3.6 for T Tau N.  The stellar contribution
is included in the models as an unresolved component.  The
visibilities show a slight dependence on hour angle, but more data are
needed to confirm this effect.

The data were reduced and binned by projected baseline before being
fit by Gaussian and uniform profile models.  The
best fit uniform profile diameter is $2.62^{+0.046}_{-0.044}$ mas (0.37
AU) and the best fit Gaussian has a FWHM of $1.61^{+0.028}_{-0.031}$
mas (0.22 AU) (Figure 1a).  For the accretion disk model, we have used
the parameters derived by \citet{ghe91} with $r_{inner}$ = 0.04 AU,
$r_{outer}$ = 100 AU, a temperature profile $T \propto r^{-0.42}$ and
$T(\rm{1~AU})$ = 260~K.  This accretion disk model overestimates the
measured visibility and, thus, underestimates the size scale.  Using
the disk parameters derived by \citet{ake98} from millimeter wave
emission ($r_{inner}$ = 0.01 AU, $r_{outer}$ = 40 AU, $T \propto
r^{-0.6}$, $T(\rm{1~AU})$ = 100~K), the predicted size is even smaller
than the Ghez model.  

\begin{figure}[ht!]
\begin{center}
\myplot{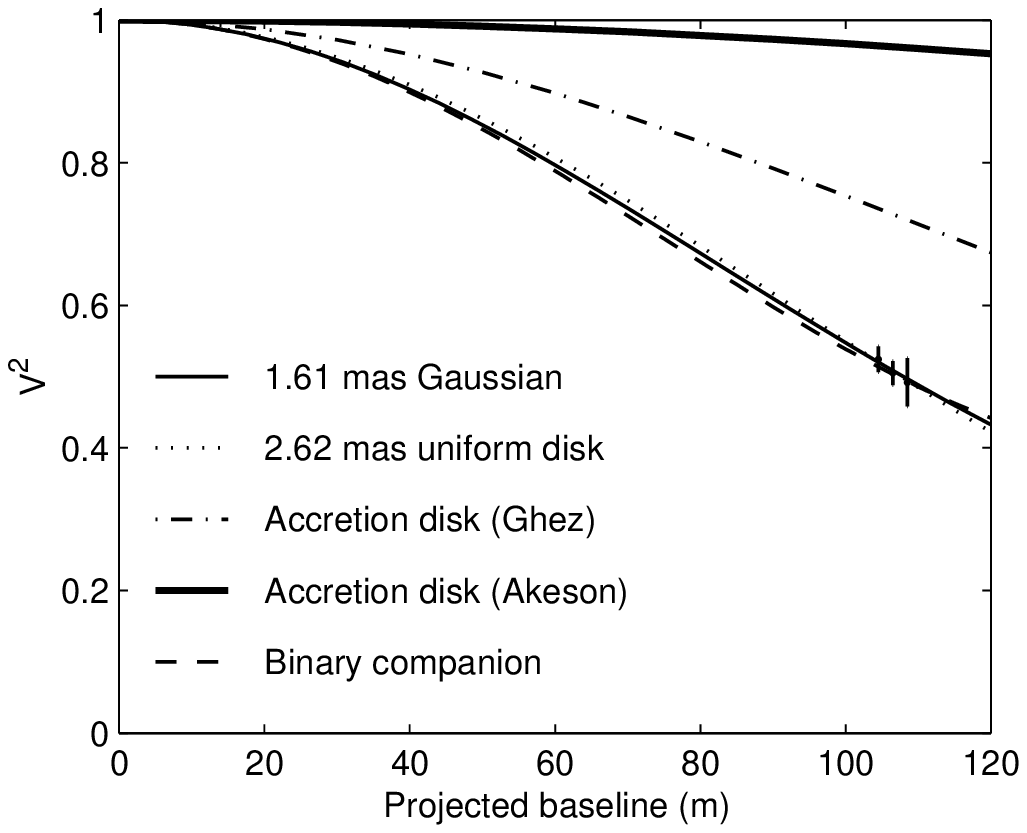}{2.2in}{0}{80}{80}{-360}{-244}
\myplot{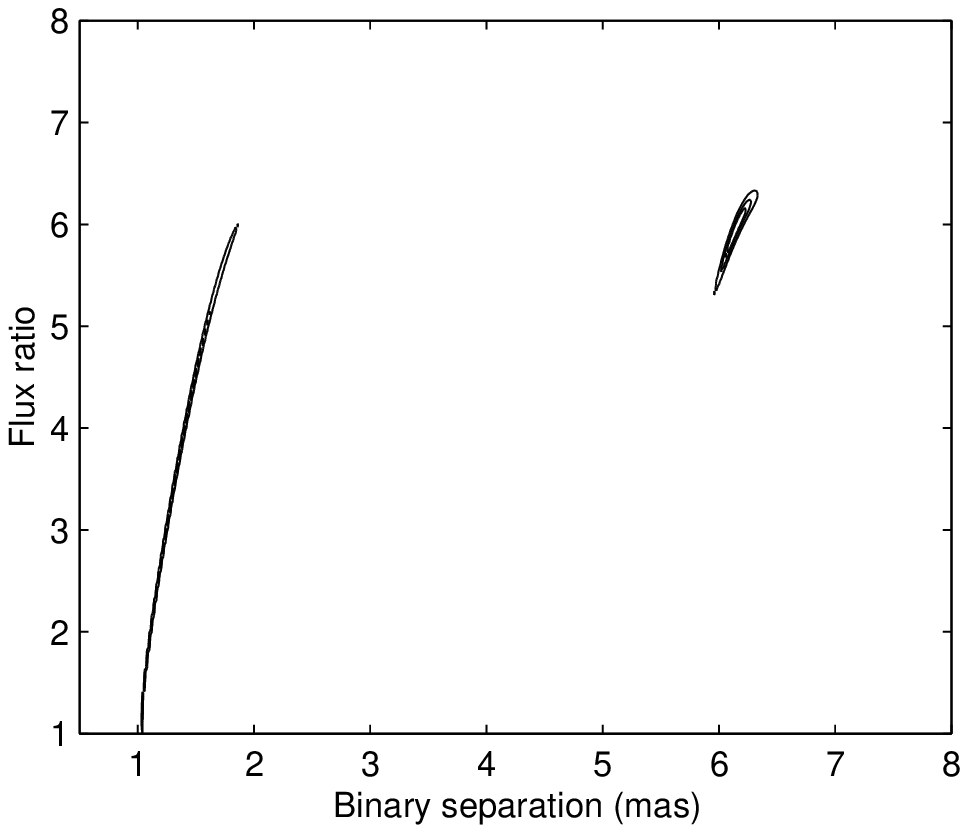}{0.0in}{0}{80}{80}{-130}{-232}
\caption{Binned data and models for T Tau.  The models are Gaussian
profile (solid line), uniform profile (dotted line), accretion disk
from \citet{ghe91} (dash-dot line), accretion disk from \citet{ake98}
(thick solid line at top) and binary companion (dashed line).  The
plotted visibilities have been corrected for the incoherent flux of T
Tau S, as described in \S \ref{ttau_sec}.  The binary parameters
represented are separation of 1.5 mas and a flux ratio of 4.3.  b)
Contour plot of possible binary companion parameters.  The contour
levels represent models with a $\chi_r^2$ of 1, 2, and 4.
\label{ttau_fig}}
\end{center}
\end{figure}

The binary parameters which fit the measured visibilities are shown in
Figure \ref{ttau_fig}b, where the contours are for $\chi_r^2$ of 1, 2
and 4.  We note that if a binary companion at the separation shown in
Figure \ref{ttau_fig}b were in a roughly circular orbit around T Tau
N, we should have seen visibility changes in the data given the time
span covered by the observations.  However, no time dependence 
was seen in the data.

\subsection{SU Aur}

\begin{figure}[ht!]
\begin{center}
\myplot{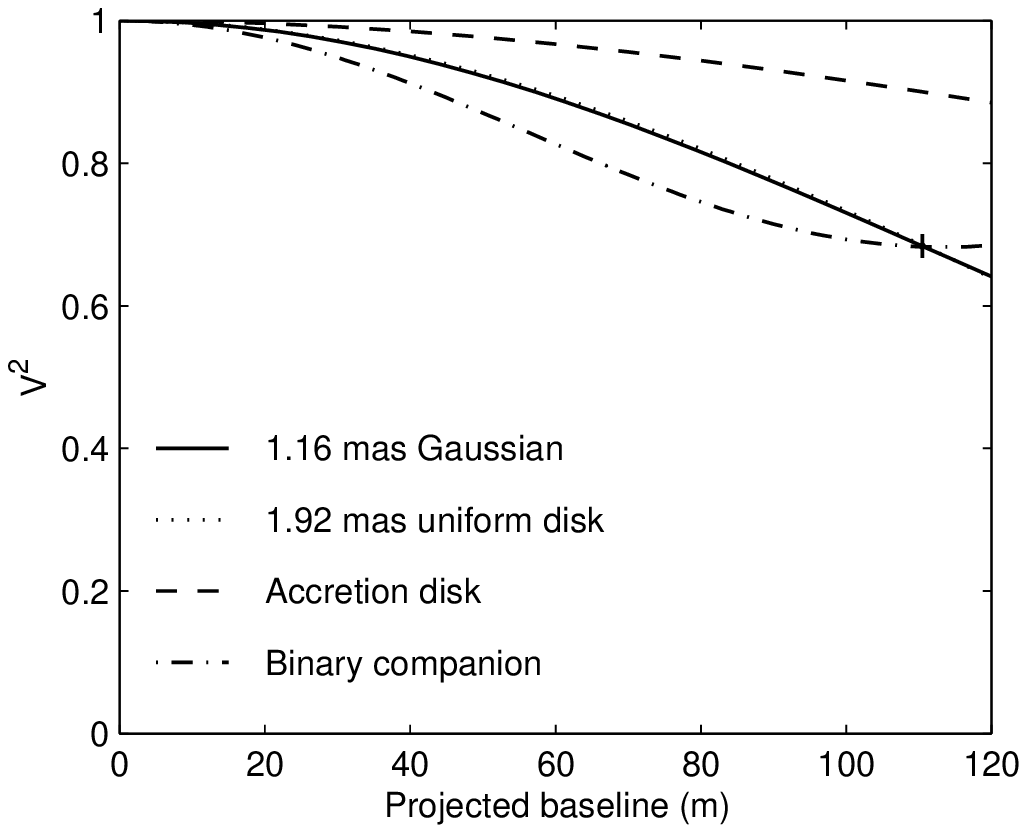}{2.2in}{0}{80}{80}{-360}{-263}
\myplot{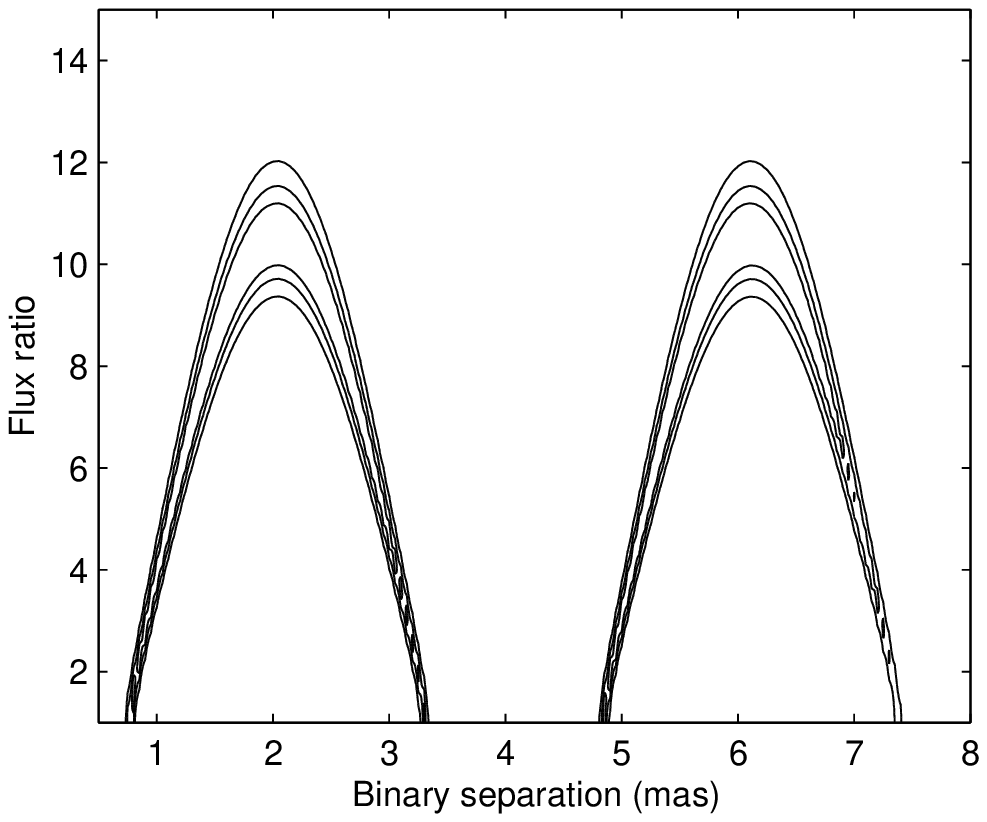}{0.0in}{0}{80}{80}{-130}{-234}
\caption{a) Binned data and models for SU Aur.  The models are
Gaussian profile (solid line), uniform profile (dotted line),
accretion disk (dashed line) and binary companion (dash-dot line). The
binary parameters represented are separation of 2 mas and a flux ratio
of 10.  b) Contour plot of possible binary companion parameters.  Note
that this pattern repeats in separation space to 20 mas.  The contour
levels represent models with a $\chi_r^2$ of 1, 2, and 4.
\label{suaur_fig}}
\end{center}
\end{figure}

SU Aur is a T Tauri star with an SED similar to that of T Tau.
\citet{her88} designated SU Aur as the prototype of a separate
classification from weak-line T Tauris due to its broad absorption
lines and high luminosity ($\sim 12~L_{\odot}$).  The stellar/total
flux ratio at K is 0.3 \citep{mar92}.  The best fit diameters are
$1.92^{+0.063}_{-0.059}$ mas and $1.16^{+0.038}_{-0.039}$ mas for a
uniform profile and a Gaussian (FWHM) respectively (Figure
\ref{suaur_fig}a).  This corresponds to a physical size of 0.27 and
0.16 AU for a distance of 140 pc.  The accretion disk model is taken
from \citet{bec90} with $r_{inner}$ = 0.01 AU, $r_{outer}$ = 100 AU, a
temperature profile $T \propto r^{-0.51}$ and $T(\rm{1~AU})$ = 260~K,
where the parameters were determined by fitting 10~$\mu$m through
millimeter wave fluxes.  This model underestimates the observed visibility.

The binary parameter space for models with reduced chi-squared,
$\chi_r^2$ of 1, 2 and 4 is shown in Figure \ref{suaur_fig}b.  The
pattern shown in the figure repeats in separation space to roughly 20
mas.  The binary companion hypothesis for SU Aur is not well
constrained due to the limited time and baseline coverage of the data.
The time coverage we do have favors orbiting companions with
separations at roughly 2 or 6 mas.

\subsection{MWC 147}

MWC~147 (HD 259431) is a Herbig Ae/Be star with spectral
classifications in the literature ranging from B2 to B6.  \citet{hil92}
modeled the SED from this source as arising from a flat,
optically thick disk with an inner hole.  Previous studies have used a
distance to this source of 800 pc, which we will use here for
consistency, although we note a recent distance determination from
Hipparcos data of $290^{+200}_{-84}$~pc \citep{ber99}.  If the
distance to MWC~147 is $\sim$290 pc, the physical sizes given below
decrease by a factor of 2.8.  Depending upon which spectral type
is used, the stellar contribution to the flux at K is 0.05 to 0.1
of the total.  We use an unresolved component with 0.1 of
the total flux to represent the central star in these models.
Using a stellar contribution of 0.05 would increase the squared
visibility due to the disk by 4\%.

\begin{figure}[ht!]
\begin{center}
\myplot{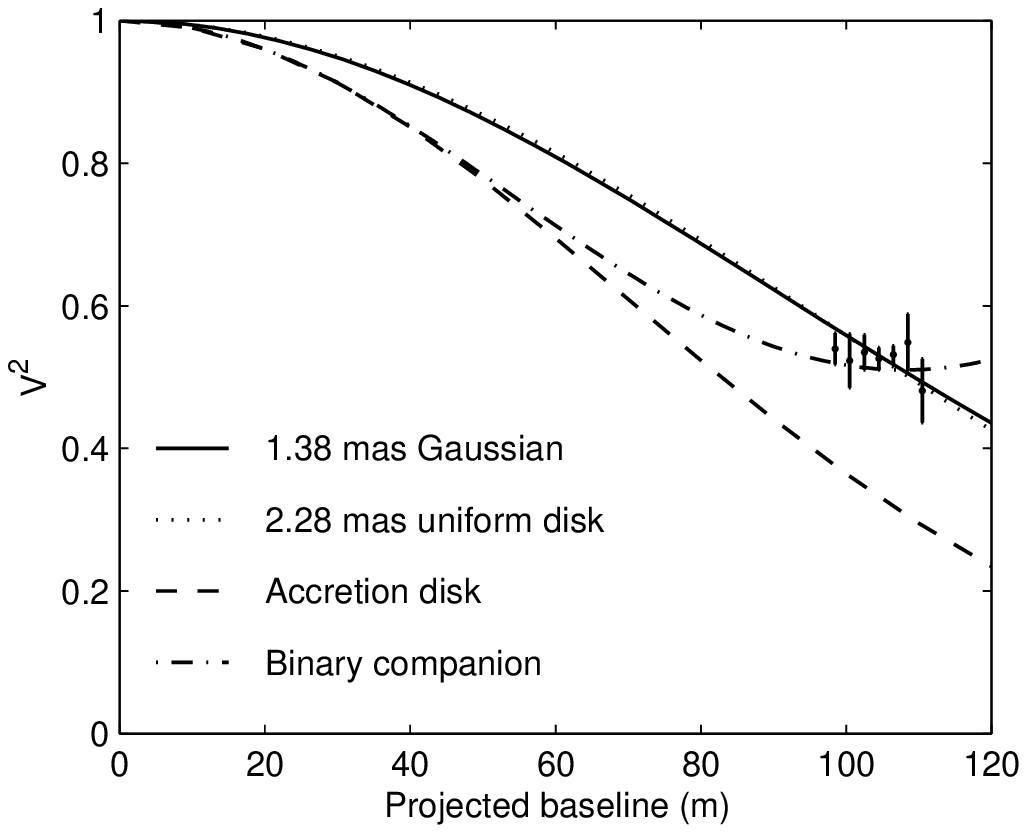}{2.2in}{0}{80}{80}{-360}{-263}
\myplot{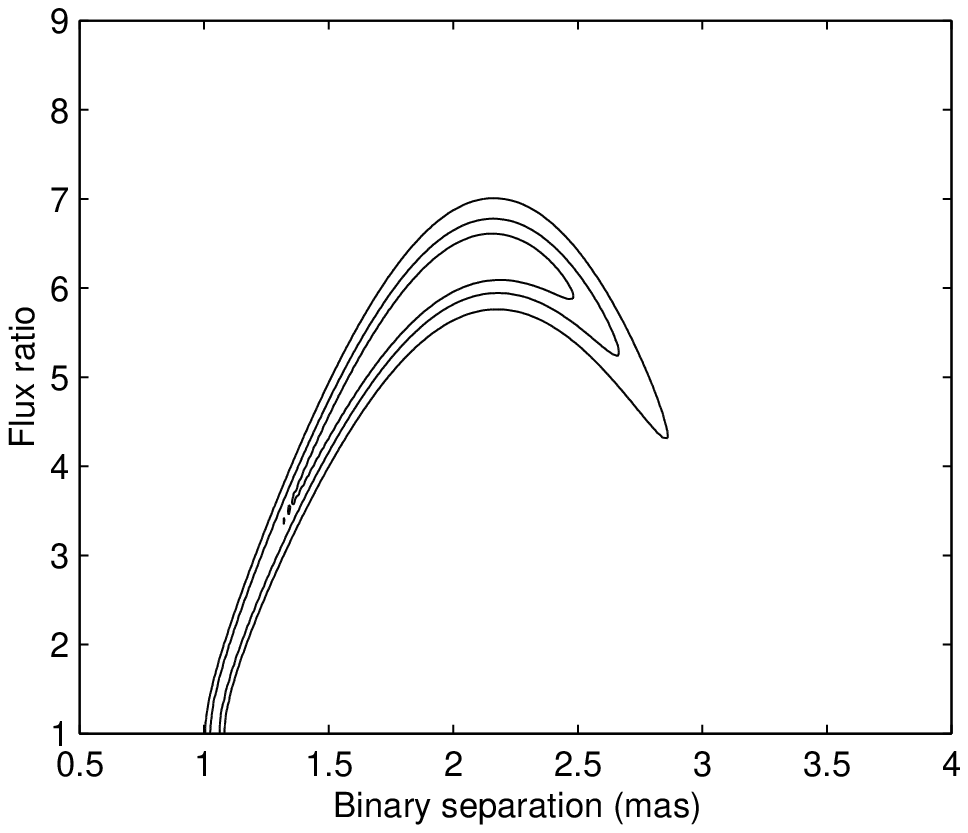}{0.0in}{0}{80}{80}{-130}{-232}
\caption{a) Binned data and models for MWC 147. The models
are Gaussian disk (solid line), uniform disk (dotted line),
accretion disk (dashed line) and binary companion
(dash-dot line).  The binary parameters represented are a separation
of 2.1 mas and a flux ratio of 6.
b) Contour plot of possible binary companion parameters.  The contour levels
represent models with a $\chi_r^2$ of 1, 2, and 4.
\label{mwc147_fig}}
\end{center}
\end{figure}

The data were reduced as described above and binned by projected
baseline.  The data and models are shown in Figure \ref{mwc147_fig}a.
Given the errors on the individual data points, there is no
significant dependence on hour angle in the visibility data,
consistent with \citet{mil99b}.  The best fit uniform profile diameter is
$2.28^{+0.017}_{-0.034}$ mas (1.8 AU) and the best fit Gaussian has a
FWHM of $1.38^{+0.013}_{-0.014}$ mas (1.1 AU).  The accretion disk and
stellar parameters were taken from \citet{hil92} with $r_{inner}$ =
0.36 AU, $r_{outer}$ = 1.8 AU and a temperature profile $T \propto
r^{-3/4}$.  The reference temperature is set by the stellar
temperature, $2 \times 10^4$~K, and the accretion rate, $\dot{M} =
10^{-5}$ M$_{\odot}$/year.  As seen in Figure \ref{mwc147_fig}a, this
accretion disk model underestimates the measured visibility and so
overestimates the physical size.

The observed visibilities for MWC 147 can also be explained by a
binary companion.  Figure \ref{mwc147_fig}b shows the parameter space
for binary models with reduced $\chi_r^2$ of 1, 2, and 4.  
An adaptive optics survey by \citet{cor98} found a
binary companion to MWC~147 with a separation of 3\farcs1 and
magnitude difference $\Delta K=5.7$.  This source is too widely
separated and too faint to have affected our observations.

\subsection{AB Aur}

AB Aur is a Herbig Ae/Be star with a spectral type of A0, at a
distance of 140 pc.  \citet{mil99a} resolved the infrared emission
from this source with the IOTA interferometer using baselines of 21
and 38 m.  At PTI, fringes were not detected on AB Aur, despite a
photon flux higher than that for T Tau, indicating that the AB Aur is
too large to be detected on baselines of $\sim$100 m with current
sensitivities.  Upper limits were found for the visibility using the
sensitivity of the detection algorithm and measuring the system
visibility with a calibrator.  At K band, the estimated upper limit
was $\langle V^2 \rangle < 0.08 \pm 0.02$, which corresponds to a size $>$4.1($\pm
0.2$) mas (0.57 AU) diameter for a uniform profile and $>$ 2.7($\pm0.1$)
mas (0.38 AU) for a Gaussian.  These results are consistent with the
size and derived by \citet{mil99a} and a previous upper limit
from PTI of $\langle V^2 \rangle < 0.3$ from \citet{ber98}.

\section{Discussion}

The fundamental result of these observations is that for all four
sources observed here, the infrared emission arising from
circumstellar material is resolved by PTI with a nominal fringe
spacing of 4 mas.  The measured sizes correspond to physical scales of
tenths of AU for the T Tauri sources T Tau and SU Aur and $\sim$1 AU
for the Herbig Ae/Be star MWC 147.  We note that if the correct
distance for MWC~147 is 290 pc, rather than 800~pc, then the measured
visibility corresponds to a size scale of roughly 0.5~AU, similar to
that measured for the T Tauri sources, despite the large difference in
stellar mass.  Our measured visibilities do not agree with those
predicted from accretion disk models derived from near-infrared SEDs
or millimeter interferometric observations.  This may suggest that the
single power-law relations used to describe the temperature and
density are inadequate to reproduce both the spectral and spatial
characteristics of the emission.

Our data on T Tau and SU Aur require the K band emission to come from
a larger region than that predicted by the accretion disk models,
while the opposite is true for MWC~147.  \citet{mil99b} observed 15
Herbig Ae/Bes using infrared interferometry with a shorter baseline
and found that roughly half of the sources could not be well modeled
as emission from an accretion disk and that the predicted visibilities
were higher than the observed data.  On the other hand, \citet{mal98}
used PTI for observations of FU Ori and found that an accretion disk
model could explain the measured visibilities.

Further characterization of the circumstellar material on size scales
less than one AU can be achieved by extending the infrared
interferometry observations presented here.  PTI has a second
baseline, which provides data on shorter spacings, and is equipped to
observe in the H band, which probes higher temperatures and has higher
spatial resolution than K band.  We plan to extend our study of
young stellar objects to include H band and more spatial scales.

\acknowledgements

This work was performed at the Infrared Processing and Analysis
Center, Caltech and the Jet Propulsion Laboratory.  Data were obtained
at the Palomar Observatory using the NASA Palomar Testbed
Interferometer, which is supported by NASA contracts to the Jet
Propulsion Laboratory.  Science operations with PTI are possible
through the efforts of the PTI Collaboration ({\tt
http://huey.jpl.nasa.gov/palomar/ptimembers.html}).  We particularly
thank A. Boden for his efforts in data reduction software and
B. Thompson for useful discussions.  We are also grateful to T. Beck
for providing the T Tau flux ratio.  DRC acknowledges support from
NASA WIRE ADP NAG5-6751.  EAL acknowledges support from a Research
Corporation Innovation Award and a Presidential Early Career Award for
Scientists and Engineers to the University of Florida.

\nopagebreak

\begin{deluxetable}{lrrrrr}
\tablecolumns{6}
\tabletypesize{\scriptsize}
\tablecaption{PTI observations}
\tablewidth{0pt}
\tablehead{
\colhead{Source} & \colhead{\# of nights} & \colhead{Calibrators} & \colhead{$<V^2>$} & \colhead{Uniform disk\tablenotemark{a}} &\colhead{Gaussian\tablenotemark{a}} \\
\colhead{} & \colhead{(records)} & \colhead{} & \colhead{} & \colhead{(diameter)} & \colhead{(FWHM)}}
\startdata
T Tau & 4(119)& HD 28024, HD 27946 & $0.29\pm0.01\tablenotemark{b}$ & $2.62^{+0.046}_{-0.044}$ mas & $1.61^{+0.028}_{-0.031}$ mas \\
SU Aur & 3(38) & HD 28024, HD 27946, HD 25867 & $0.68\pm0.02$ & $1.92^{+0.063}_{-0.059}$ mas& $1.16^{+0.038}_{-0.039}$ mas \\
MWC 147 & 9(121) & HD 43042, HD 46709 & $0.53\pm0.01$ & $2.28^{+0.017}_{-0.034}$ mas & $1.38^{+0.013}_{-0.014}$ mas \\
AB Aur & 4\tablenotemark{c} & HD 32301 & $<0.15$ & $>$3.7 mas & $>$2.4 mas \\ \tableline
\enddata
\tablenotetext{a}{1 $\sigma$ uncertainties are given for the best fit
Gaussian and uniform disk models.}
\tablenotetext{b}{The visibility given for T Tau is the calibrated value
uncorrected for the effects of T Tau S.}
\tablenotetext{c}{Only nights with good upper limits are listed
for AB Aur}
\end{deluxetable}

\end{document}